\documentclass{article}
\usepackage{spconf,amsmath,epsfig}

\def\x{{\mathbf x}}

\def\s{{\mathbf s}}

\def\R{{\mathbf R}}
\def\w{{\mathbf w}}
\def\I{{\mathbf I}}
\def\X{{\mathbf X}}

\def\v{{\mathbf v}}

\newtheorem{proof sketch}{Proof Sketch}

\usepackage{setspace}
\usepackage{cite}
\usepackage{graphicx}
\usepackage{subfigure}
\usepackage{amsmath}
\usepackage{psfig}
\usepackage{psfrag}
\usepackage{amssymb}
\usepackage{paralist}

\graphicspath{{Figures/}}

\title{Optimization of loading factor preventing target cancellation}
\name{Boris N. Oreshkin$^\dag$ and Peter A. Bakulev$^\ddag$}
\address{$^\dag$Department of Electrical and Computer Engineering, McGill University\\
$^\ddag$School of Aircraft Electronics, Moscow State Aviation
Institute}

\begin{document}
%
\maketitle
\begin{abstract}
Adaptive algorithms based on sample matrix inversion belong to an
important class of algorithms used in radar target detection to
overcome prior uncertainty of interference covariance. Sample matrix
inversion problem is generally ill conditioned. Moreover, the
contamination of the empirical covariance matrix by the useful
signal leads to significant degradation of performance of this class
of adaptive algorithms. Regularization, also known in radar
literature as sample covariance loading, can be used to combat both
ill conditioning of the original problem and contamination of the
empirical covariance by the desired signal. However, the optimum
value of loading factor cannot be derived unless strong assumptions
are made regarding the structure of covariance matrix and useful
signal penetration model. In this paper an iterative algorithm for
loading factor optimization based on the maximization of empirical
signal to interference plus noise ratio (SINR) is proposed. The
proposed solution does not rely on any assumptions regarding the
structure of empirical covariance matrix and signal penetration
model. The paper also presents simulation examples showing the
effectiveness of the proposed solution.
\end{abstract}
\begin{keywords}
adaptive filters, matrix inversion, interference suppression.
\end{keywords}
%
\section{Introduction}
\label{section::Introduction} 
Loading of the sample covariance matrix is used in loaded sample
matrix inversion (LSMI) algorithm to alleviate losses incurred by
using finite sample size estimate of the true covariance
matrix~\cite{Car88}. The well known heuristic result due to
Carlson~\cite{Car88} suggests fixing the value of loading factor at
the level 10 dB above the white noise power. This choice of loading
factor is indeed suitable in many practical scenarios. However, in
some situations this value of loading factor might turn out to be
too low, \emph{e.g.}, when useful signal is present in the training
sample that is used for covariance matrix estimation. In this
situation the effectiveness of LSMI using fixed loading factor
significantly decreases. Moreover, if the signal power in training
sample is sufficiently strong, target cancellation may occur.
According to the well established methodology, loading factor can be
optimized to prevent target cancellation by taking into account
phased array calibration errors~\cite{Bec07,Vor03}. This approach
relies on a reasonably accurate model for training sample
contamination. Often this model does not reflect the actual
mechanisms that lead to the contamination of the training sample by
the useful signal. For example, when the target is highly
manoeuvring or distributed, this approach cannot model useful signal
penetration. In this paper we concentrate on the non--parametric
approach to loading factor optimization that does not rely on any
model for training sample contamination. This approach is based on
the iterative maximization of empirical Rayleigh quotient.

The remaining of this paper is organized as follows.
Section~\ref{section:Problem Statement} describes the LSMI algorithm
and presents loading factor optimization problem.
Section~\ref{section:load_fact_optim} describes the solution to the
optimization problem and proposed iterative algorithm for the
optimization of loading factor. Section~\ref{section:num_examp}
presents numerical examples showing the effectiveness of proposed
algorithm. Finally, section~\ref{section::Concluding Remarks}
concludes the paper.

\section{Problem Statement}
\label{section:Problem Statement}
The problem of finding optimum weights for the linear detector of a
known signal $\s$ corrupted by the correlated interference with
known covariance $\R$ can be formulated in terms of the output SINR
maximization \cite{Vor03}:
\begin{equation} \label{eqn:lsmi_Ray}
\gamma_\text{out} = \frac{|\w^H \s|^2}{\w^H \R \w}.
\end{equation}
Here $\gamma_\text{out}$ is the output SINR of the detector, also
known as Rayleigh quotient, and $\w$ are detector weights. The
expression for $\w$ maximizing (\ref{eqn:lsmi_Ray}) is known to have
the following form~\cite{Car88}:
\begin{equation} \label{eqn:w}
\w_\text{opt} \propto \R^{-1} \s.
\end{equation}
When $\R$ is not known in advance, one can resort to the adaptive
version of (\ref{eqn:w}) using the maximum likelihood estimate of
covariance matrix~\cite{Car88}:
\begin{equation} \label{eqn:ML_R}
\widehat \R = \frac{1}{M} \X \X^H,
\end{equation}
instead of the true $\R$. Here $\X = [\x_1, \x_2, \ldots, \x_M]$ is
a training sample containing $N \times 1$ vectors of the space--time
samples $\{ \x_i \}_{i=1}^M$ of locally homogeneous correlated
Gaussian interference derived from a set of cells surrounding the
cell of interest, $N$ is the dimensionality of space--time
processing and $M$ is the number of training samples. The
introduction of loading factor $\alpha$ leads to the following
expression for the regularized estimate of the interference
covariance matrix:
\begin{equation} \label{eqn:reg_R_hat}
\widehat \R^{\prime} = \widehat \R + \alpha \I,
\end{equation}
leading to the adaptive LSMI algorithm:
\begin{equation} \label{eqn:lsmi_w}
\widehat \w \triangleq \w_\text{LSMI} = \left(\widehat \R + \alpha\I
\right)^{-1}\s.
\end{equation}
Given the structure (\ref{eqn:lsmi_w}) imposed on the estimator of
optimum weights and the fact that optimum $\alpha$ is unknown, the
following optimization problem can be stated:
\begin{equation} \label{eqn:lsmi_const_opt}
\widehat \alpha_\text{opt} = \arg\max\limits_{\alpha} \widehat
\gamma_\text{out}
\end{equation}
Here we introduce $\widehat \gamma_\text{out}$ to clarify our
assumption that $\widehat \w$ defined in (\ref{eqn:lsmi_w}) and
$\widehat \R^{\prime}$ defined in (\ref{eqn:reg_R_hat}) can be used
to approximate $\gamma_\text{out}$ by the empirical SINR $\widehat
\gamma_\text{out}$ while maximizing $\gamma_\text{out}$ with respect
to $\alpha$.

\section{Loading Factor Optimization}
\label{section:load_fact_optim}
In the following we consider solving the problem
(\ref{eqn:lsmi_const_opt}). At first, we use the expression
(\ref{eqn:lsmi_Ray}) for the true $\gamma_\text{out}$ to reformulate
the problem. After that, we use the necessary approximations to
arrive at the expression for the cost function.

\begin{figure}[t]
\centering
\begin{tabular}{|l|}
\hline ~\\
\begin{minipage}[l]{8cm}
\begin{enumerate}

\item Covariance matrix estimation\\

\vspace{-0.3cm} $\widehat \R = \frac{1}{M} \X \X^H$ \ \ \
(\ref{eqn:ML_R});

\item Initialization of iterative algorithm\\

\vspace{-0.3cm} $\alpha_1 = \sigma^2_n$;

\item For all $i = 1\ldots T$

$\widetilde \w_i = \left(\widehat\R + \alpha_i\I \right)^{-1}\s$   \
\ \ (\ref{eqn:lsmi_not}); \\

$\widetilde\v_i = \left(\widehat\R + \alpha_i\I
\right)^{-1}\widetilde\w_i$ \ \ \ (\ref{eqn:lsmi_not}); \\

$\lambda_i = -1 - \frac{{2\widetilde\w_i^H\widetilde\v_i(1 -
\widetilde\w_i^H\s) +
\widetilde\w_i^H\widetilde\w_i\widetilde\v_i^H\s}}{{|\s^H\widetilde\v_i|^2}}$
\ \ \ (\ref{eqn:lsmi_lam2}); \\

$\alpha_{i+1} = \frac{{\widetilde\w_i^H\widetilde\w_i +
\widetilde\v_i^H\s +
\lambda_i\widetilde\v_i^H\s}}{({2\widetilde\v_i^H\widetilde\w_i})}$
\ \ \ (\ref{eqn:lsmi_alpha});
\item Estimation of weight vector\\

\vspace{-0.3cm} $\widehat\w = \left(\widehat\R + \alpha_{T}\I
\right)^{-1}\s$ \ \ \ (\ref{eqn:lsmi_w});

\end{enumerate}
\end{minipage}
~\\~\\\hline
\end{tabular}
\caption{Iterative algorithm for estimation of loading factor
$\alpha$ and weight vector $\widehat\w$.} \label{fig:it_rokm}
\vspace{0.5cm}
\end{figure}

To facilitate algebraic manipulations, the problem of maximizing
$\gamma_\text{out}$ in (\ref{eqn:lsmi_Ray}) with respect to $\alpha$
can be reformulated in terms of minimizing $\w^H \R \w$ subject to a
constraint on $|\w^H \s|^2$:
\begin{equation} \label{eqn:alpha_opt_reform}
\alpha_\text{opt} = \arg\min\limits_{\alpha} \w^H \R \w, \text{\ \
s. t.\ \ } |\w^H \s|^2 = 1
\end{equation}
To resolve (\ref{eqn:alpha_opt_reform}) we apply approximations $\w
\simeq \widehat\w$ and $\R \simeq \widehat\R^\prime$ to $\w$ and
$\R$ in (\ref{eqn:lsmi_Ray}) and use the linearization of
(\ref{eqn:lsmi_w}) proposed in~\cite{Tia01}. Note that $\alpha$ is
small relative to the diagonal entries of $\widehat \R$ and we can
use Taylor series to approximate (\ref{eqn:lsmi_w}) in the vicinity
of the point $\alpha = 0$:
\begin{equation} \label{eqn:lsmi_w_ser}
\begin{split}
\widehat\w &= \left[\widehat\w\right]_{\alpha = 0} + \alpha \left[
\frac{\partial}{\partial \alpha} \widehat\w \right]_{\alpha = 0} +
\frac{\alpha^2}{2!} \left[ \frac{\partial^2}{\partial\alpha^2}
\widehat\w \right]_{\alpha = 0}\\
&+ \frac{\alpha^3}{3!} \left[ \frac{\partial^3}{\partial\alpha^3}
\widehat\w \right]_{\alpha = 0} + \ldots
\end{split}
\end{equation}
Calculating the first derivative of $\widehat\w$ with respect to
$\alpha$:
\begin{equation} \label{eqn:lsmi_w_1st_dir}
\begin{split}
\left[ \frac{\partial}{\partial \alpha} \widehat\w \right]_{\alpha =
0} &= \left[ -\left(\widehat \R + \alpha\I
\right)^{-1}\I\left(\widehat \R
+ \alpha\I \right)^{-1}\s \right]_{\alpha = 0}\\
&= -\widehat \R^{-2}\s,
\end{split}
\end{equation}
and limiting the expansion (\ref{eqn:lsmi_w_ser}) to its first two
components results in the following linear approximation of
(\ref{eqn:lsmi_w}):
\begin{equation} \label{eqn:lsmi_w_approx}
\widehat\w \simeq \widehat \R^{-1}\s - \alpha \widehat \R^{-2}\s.
\end{equation}
The introduction of the notation
\begin{equation} \label{eqn:lsmi_not}
\widetilde \w \triangleq \widehat\R^{-1}\s \text{\ and \ }
\widetilde \v \triangleq \widehat\R^{-1}\widetilde\w
\end{equation}
leads to the representation of (\ref{eqn:lsmi_w_approx}) in terms of
$\widetilde \w$ and $\widetilde \v$
\begin{equation} \label{eqn:lsmi_w_not_approx}
\widehat\w \simeq \widetilde\w - \alpha\widetilde\v.
\end{equation}
Substituting (\ref{eqn:lsmi_w_not_approx}) into
(\ref{eqn:alpha_opt_reform}) and using Lagrange multipliers results
in the following cost function to be minimized:
\begin{equation} \label{eqn:lsmi_J}
\begin{split}
J(\alpha) &= \widetilde\w^H\s - \alpha \widetilde\w^H\widetilde\w
-\alpha\widetilde\v^H\s\\
&+ \alpha^2\widetilde\v^H\widetilde\w + \lambda(\widetilde\w^H\s -
\alpha\widetilde\v^H\s - 1),
\end{split}
\end{equation}
where $\lambda$ is the Lagrange multiplier. The derivative of the
cost function with respect to $\alpha$ has the following expression:
\begin{equation} \label{eqn:lsmi_J_grad}
\frac{\partial}{\partial\alpha}J(\alpha) =
-\widetilde\w^H\widetilde\w - \widetilde\v^H\s +
2\alpha\widetilde\v^H\widetilde\w - \lambda\widetilde\v^H\s.
\end{equation}
Equating (\ref{eqn:lsmi_J_grad}) to zero we find the expression for
$\widehat\alpha_\text{opt}$:
\begin{equation} \label{eqn:lsmi_alpha}
\widehat\alpha_\text{opt} = \frac{\widetilde\w^H\widetilde\w +
\widetilde\v^H\s +
\lambda\widetilde\v^H\s}{2\widetilde\v^H\widetilde\w}.
\end{equation}
\begin{figure*}[tbp]
\centering \subfigure[Input SINR = 20 dB]{
\label{fig:reg_aut_3:20} 
\includegraphics[width = 5cm]{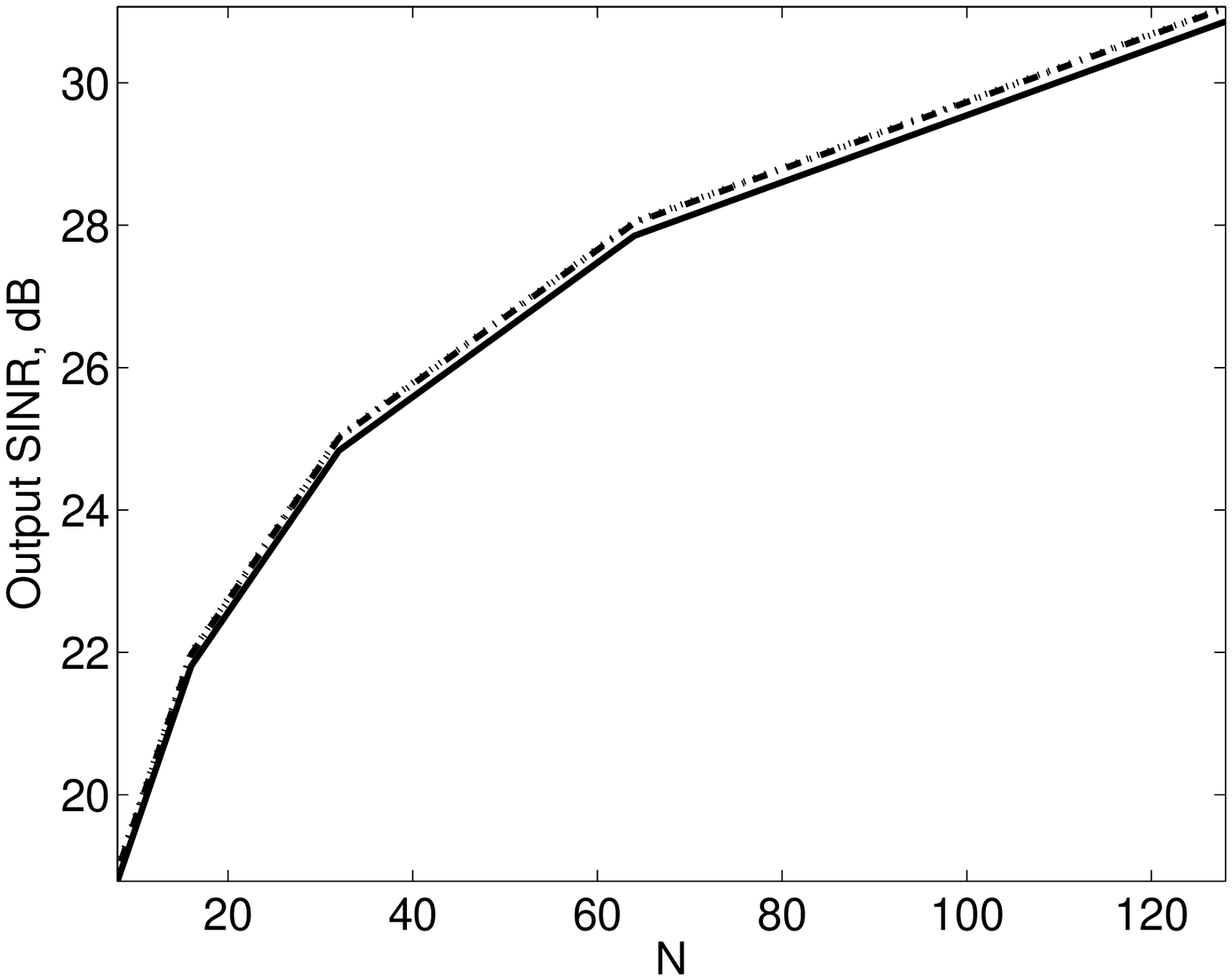}}
\hfill \subfigure[Input SINR = 0 dB]{
\label{fig:reg_aut_3:0} 
\includegraphics[width = 5cm]{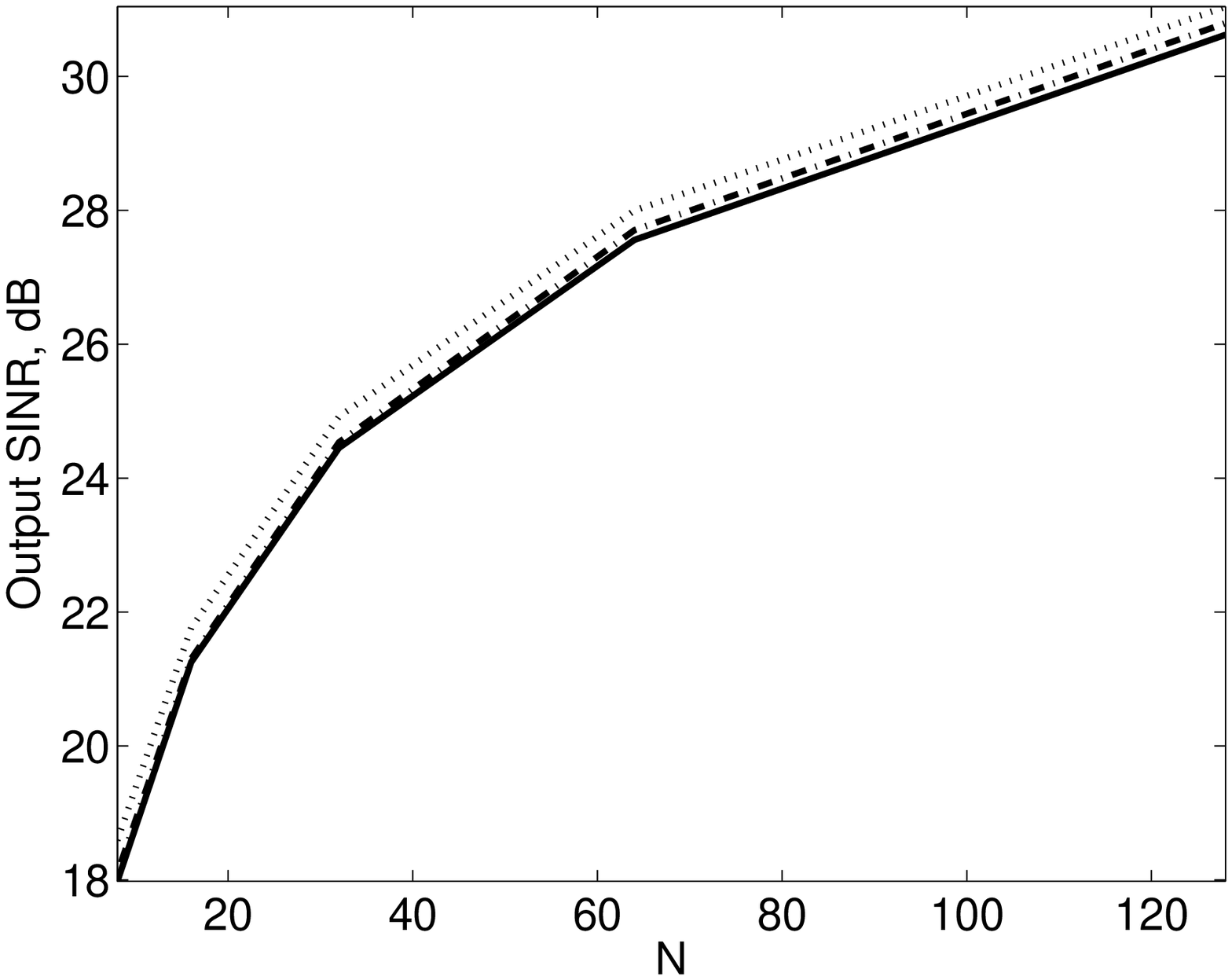}}
\hfill \subfigure[Input SINR = -20 dB]{
\label{fig:reg_aut_3:-20} 
\includegraphics[width = 5cm]{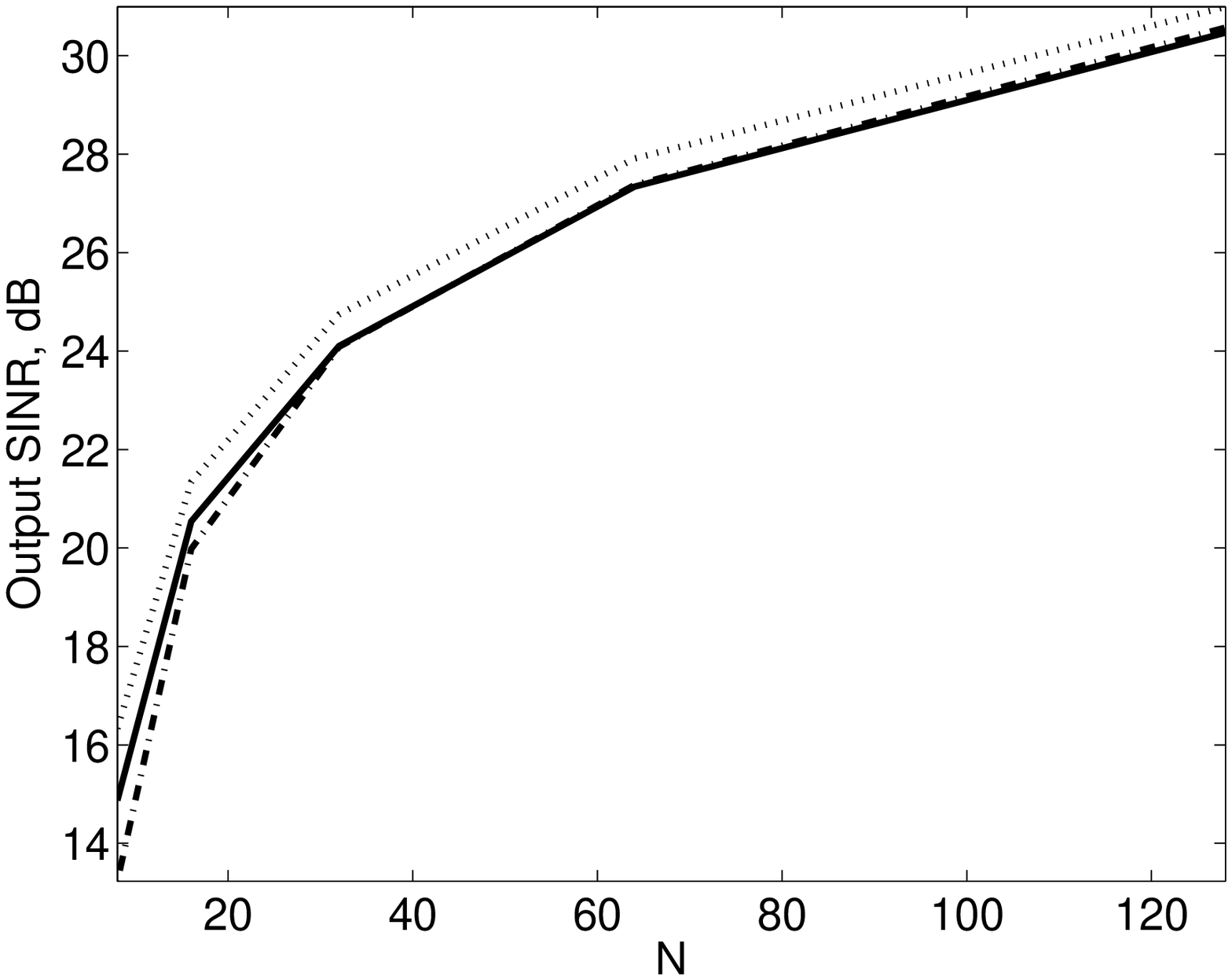}}
\hfill \subfigure[Input SINR = -40 dB]{
\label{fig:reg_aut_3:-40} 
\includegraphics[width = 5cm]{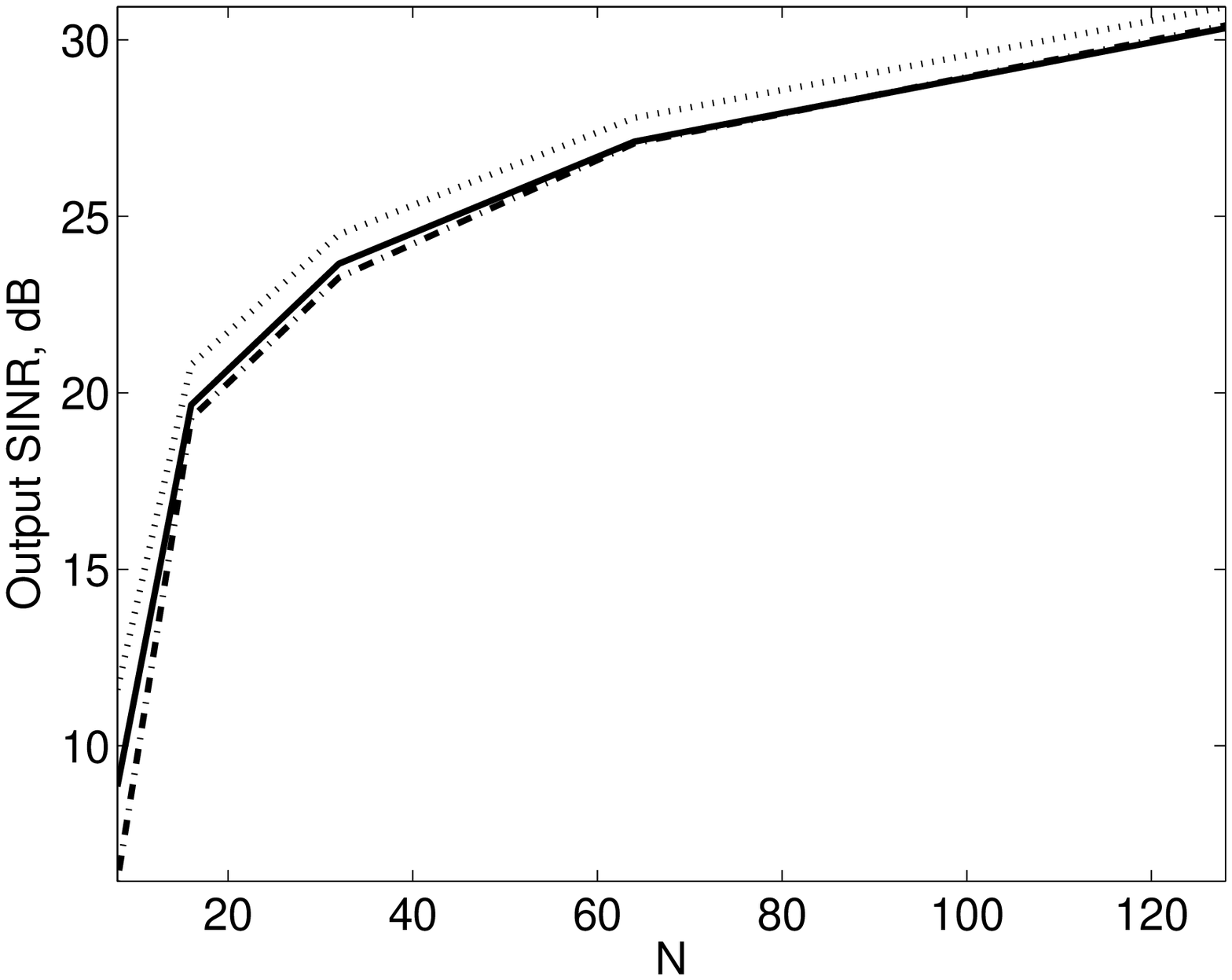}}
\hfill \subfigure[Input SINR = -60 dB]{
\label{fig:reg_aut_3:-60} 
\includegraphics[width = 5cm]{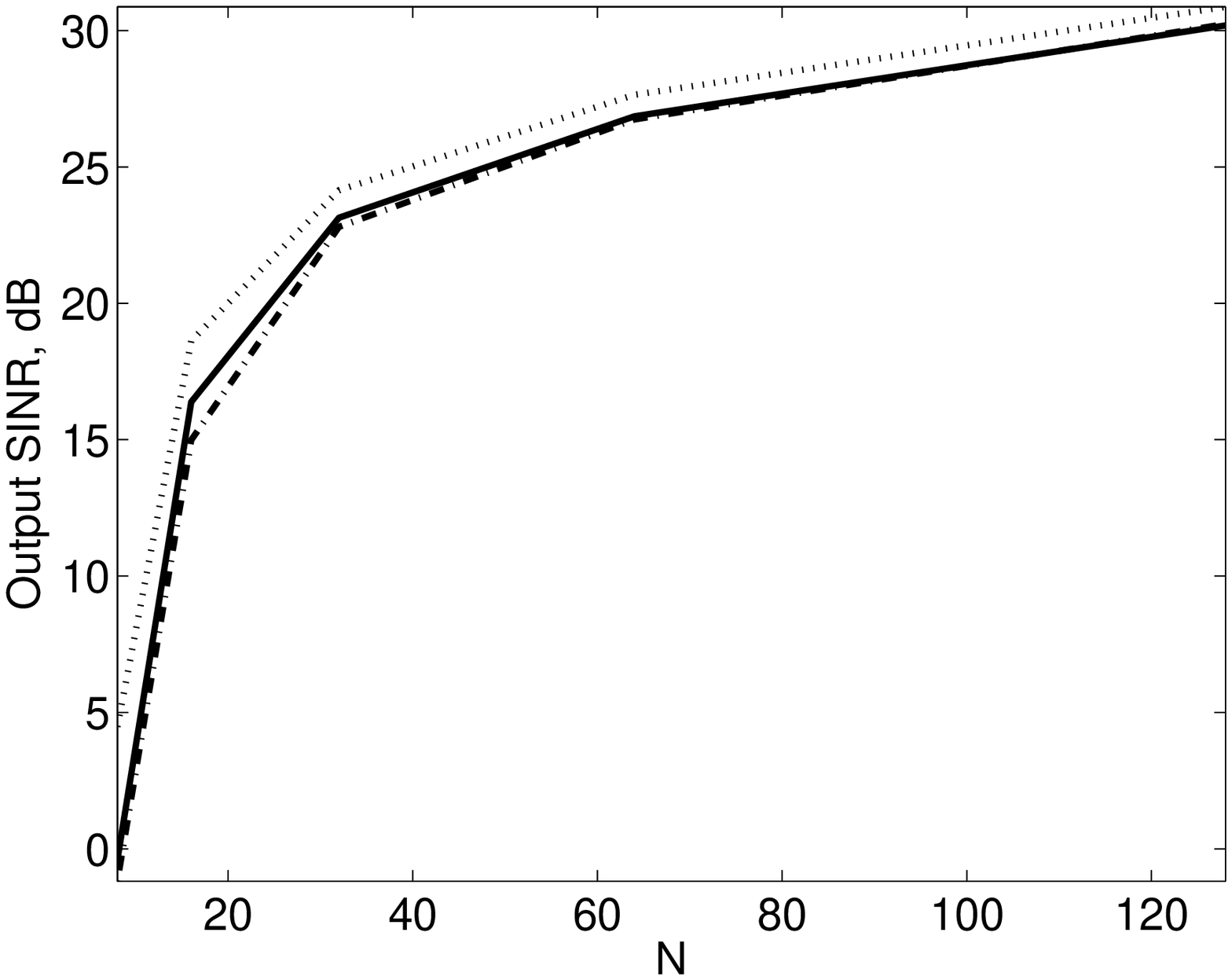}}
\hfill \subfigure[Input SINR = -80 dB]{
\label{fig:reg_aut_3:-80} 
\includegraphics[width = 5cm]{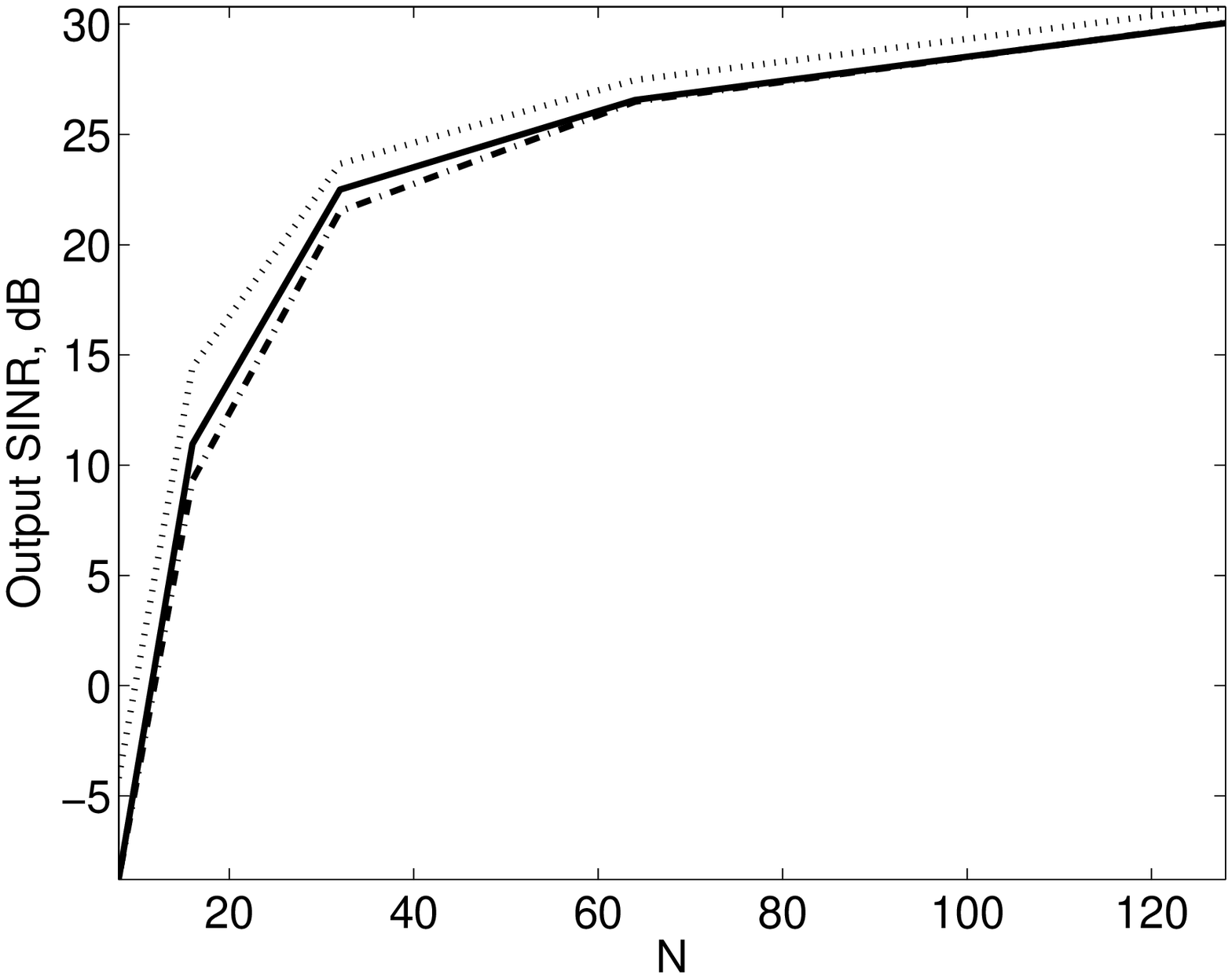}}
\caption{Output SINR of LSMI as a function of the number of samples
$N$ for different values of input SINR. Dotted line denotes optimum
algorithm with known covariance $\R$, solid line denotes proposed
approach, dash--dotted line denotes LSMI with fixed loading factor
$\alpha = 10\sigma_n^2$~\cite{Car88}. Training sample is not
contaminated by the useful signal, $M = N$.}
\label{fig:reg_aut_1} 
\end{figure*}
\begin{figure*}[tbp]
\centering \subfigure[Input SINR = 20 dB]{
\label{fig:reg_aut_4:20} 
\includegraphics[width = 5cm]{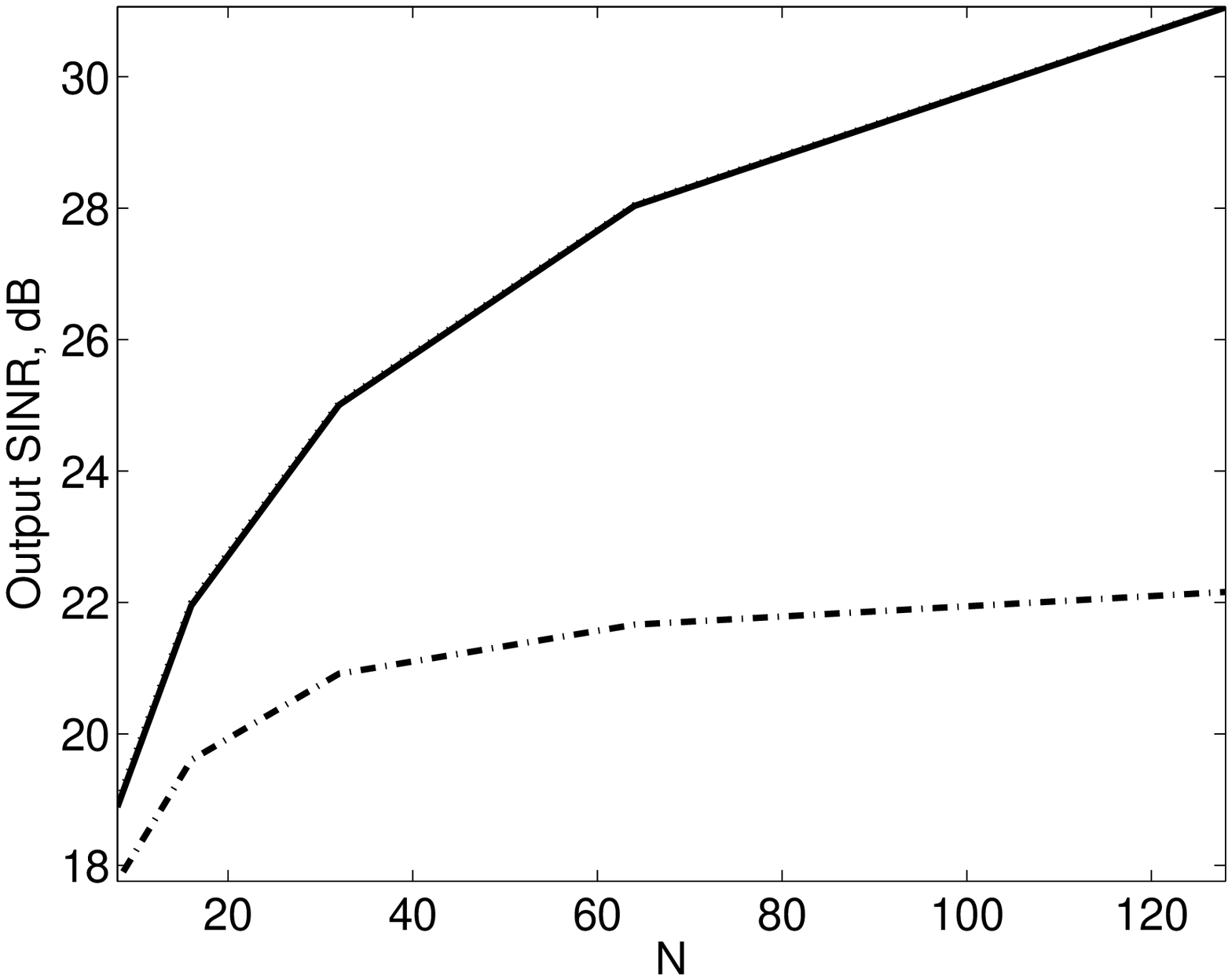}}
\hfill \subfigure[Input SINR = 0 dB]{
\label{fig:reg_aut_4:0} 
\includegraphics[width = 5cm]{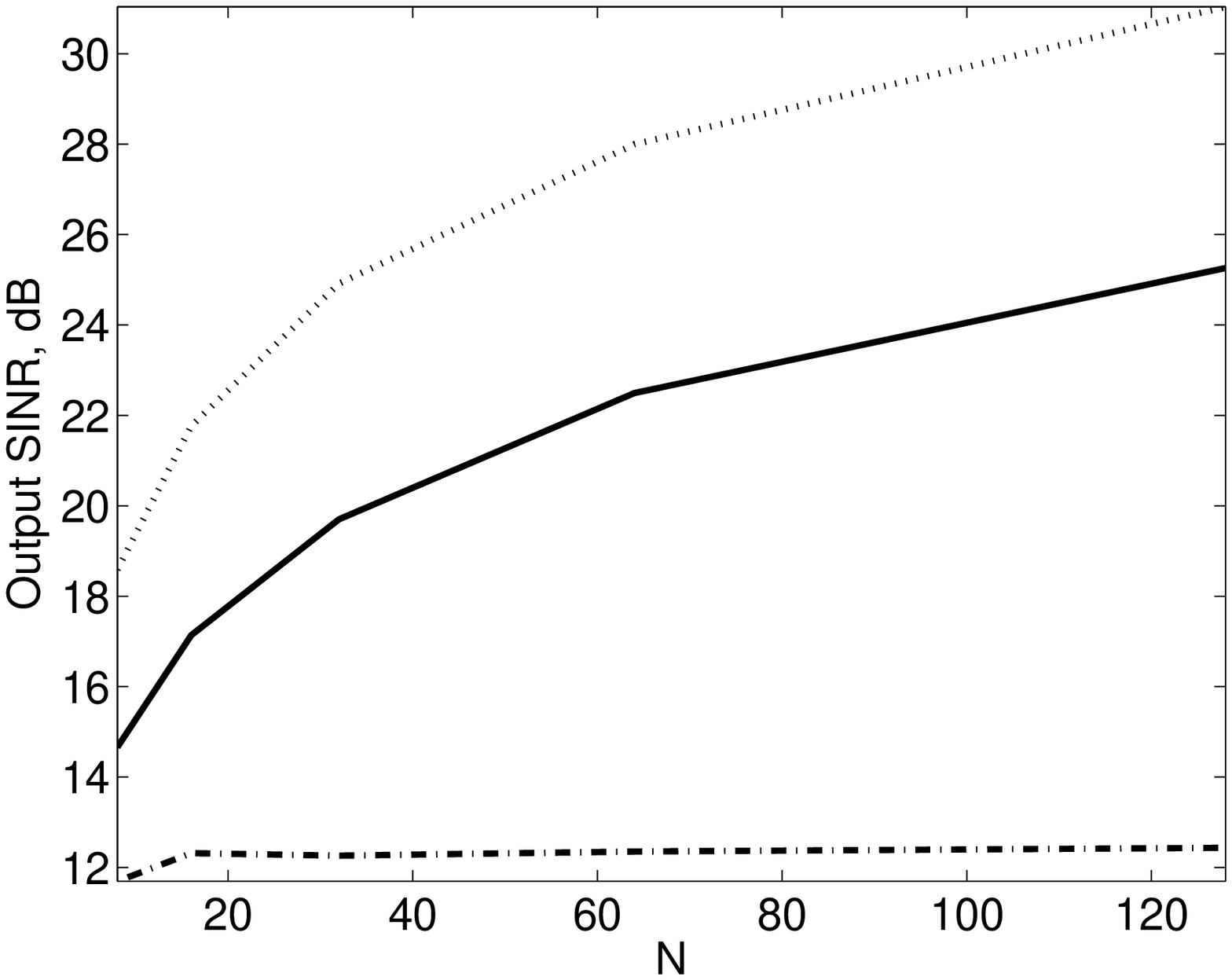}}
\hfill \subfigure[Input SINR = -20 dB]{
\label{fig:reg_aut_4:-20} 
\includegraphics[width = 5cm]{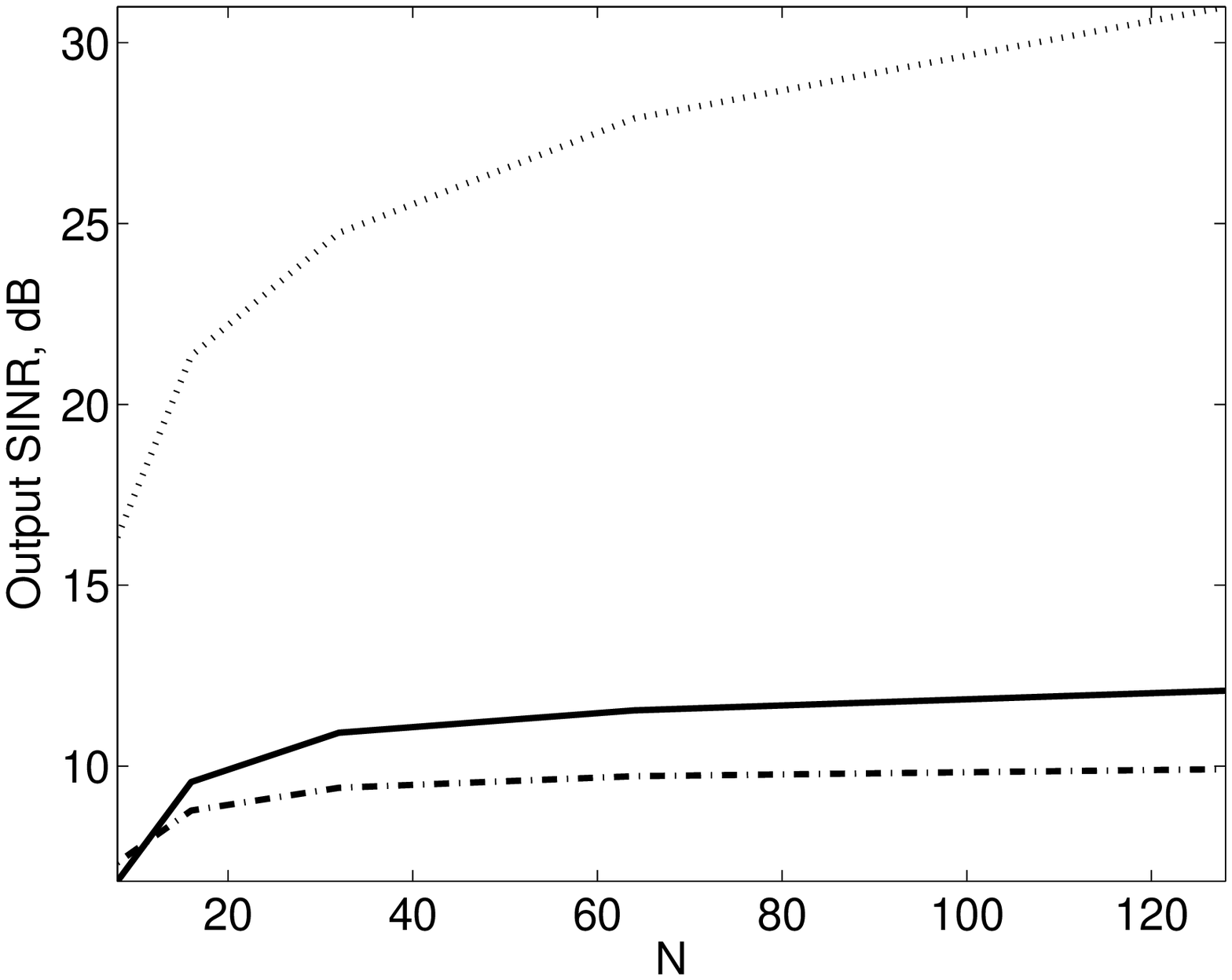}}
\hfill \subfigure[Input SINR = -40 dB]{
\label{fig:reg_aut_4:-40} 
\includegraphics[width = 5cm]{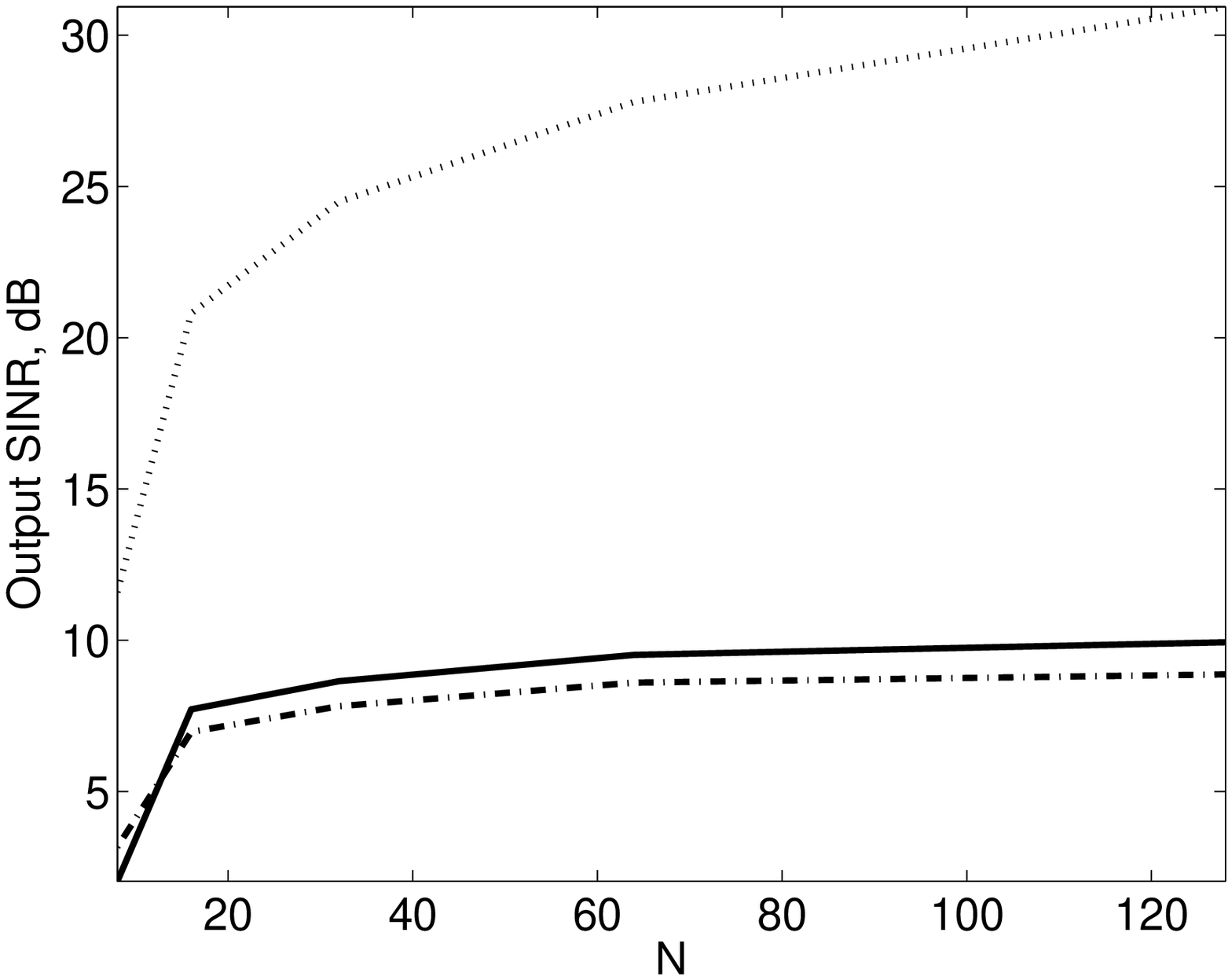}}
\hfill \subfigure[Input SINR = -60 dB]{
\label{fig:reg_aut_4:-60} 
\includegraphics[width = 5cm]{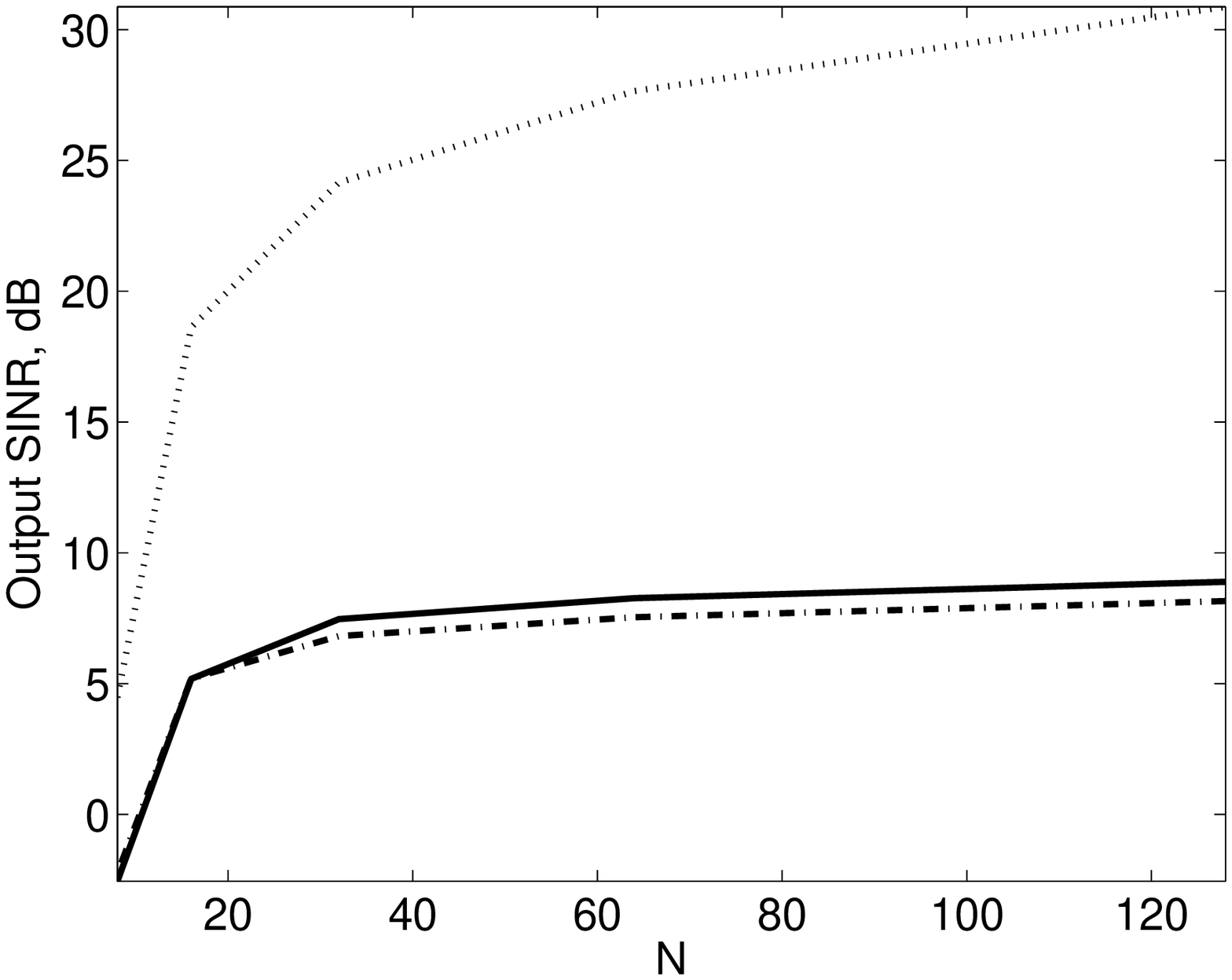}}
\hfill \subfigure[Input SINR = -80 dB]{
\label{fig:reg_aut_4:-80} 
\includegraphics[width = 5cm]{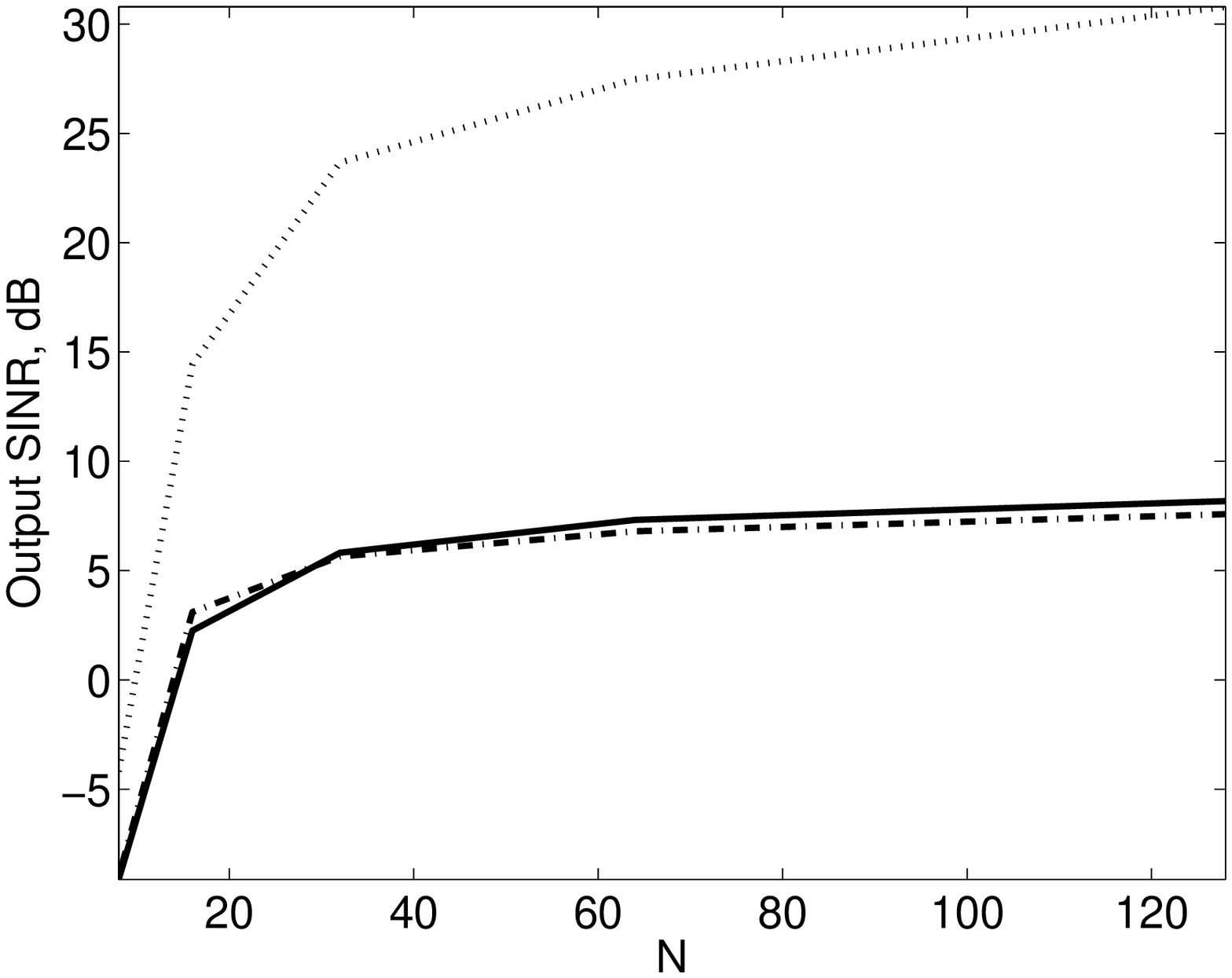}}
\caption{Output SINR of LSMI as a function of the number of samples
$N$ for different values of input SINR. Dotted line denotes optimum
algorithm with known covariance $\R$, solid line denotes proposed
approach, dash--dotted line denotes LSMI with fixed loading factor
$\alpha = 10\sigma_n^2$~\cite{Car88}. Training sample is
contaminated by the useful signal, $M = N$.}
\label{fig:reg_aut_2} 
\end{figure*}
The expression for the unknown $\lambda$ follows from substituting
(\ref{eqn:lsmi_alpha}) into constraint equation $|\w^H \s|^2 = 1$
following from (\ref{eqn:alpha_opt_reform}):
\begin{equation} \label{eqn:lsmi_lam1}
\begin{split}
\widehat\w^H\s &= (\widetilde\w^H - \widehat\alpha_\text{opt}^*\widetilde\v^H)\s \\
&= \widetilde\w^H\s - \left( \frac{\widetilde\w^H\widetilde\w +
\widetilde\v^H\s +
\lambda\widetilde\v^H\s}{2\widetilde\v^H\widetilde\w} \right)^{*}\widetilde\v^H\s  \\
& = \widetilde\w^H\s -
\frac{\widetilde\w^H\widetilde\w\widetilde\v^H\s +
\s^H\widetilde\v\widetilde\v^H\s +
\lambda\s^H\widetilde\v\widetilde\v^H\s}{2\widetilde\w^H\widetilde\v}
= 1 \nonumber
\end{split}
\end{equation}
Finally, the expression for $\lambda$ becomes:
\begin{equation} \label{eqn:lsmi_lam2}
\lambda = -1 - \frac{2\widetilde\w^H\widetilde\v(1 -
\widetilde\w^H\s) +
\widetilde\w^H\widetilde\w\widetilde\v^H\s}{|\s^H\widetilde\v|^2}
\end{equation}
Thus the algorithm for data dependent $\alpha$ optimizaton,
maximizing empirical SINR $\widehat \gamma_\text{out}$ consists of
calculating $\widetilde\w$ and $\widetilde\v$ using
(\ref{eqn:lsmi_not}) and $\lambda$ and $\widehat \alpha_\text{opt}$
using (\ref{eqn:lsmi_lam2}) and (\ref{eqn:lsmi_alpha}). Finally,
$\widehat\w$ is calculated according to (\ref{eqn:lsmi_w}). Linear
approximation in (\ref{eqn:lsmi_w_approx}) prevents the convergence
of the algorithm outlined above to $\widehat \alpha_\text{opt}$
during the first iteration if $\widehat \alpha_\text{opt}$ is
comparatively large. We propose to use iterative algorithm to
converge to this value. This algorithm is outlined in
Fig.~\ref{fig:it_rokm}. Note that this algorithm is initialized with
the value of $\alpha_1$ equal to the power of white noise
$\sigma_n^2$. The idea behind this iterative algorithm is that at
$i$--th iteration of the algorithm, $\widehat \w$ is expanded around
the point $\alpha = \alpha_i$ and as $\alpha_i$ becomes closer to
the true $\widehat \alpha_\text{opt}$ with growing $i$, the accuracy
of approximation (\ref{eqn:lsmi_w_not_approx}) increases leading to
the convergence of the algorithm to $\widehat \alpha_\text{opt}$.

\section{Numerical Examples}
\label{section:num_examp}
In this section we compare the performance of optimum algorithm with
known covariance $\R$, LSMI with fixed loading factor $\alpha =
10\sigma_n^2$~\cite{Car88}, and proposed approach to data--dependent
estimation of $\alpha$ in terms of output SINR.
Figures~\ref{fig:reg_aut_1} and~\ref{fig:reg_aut_2} show the results
of this comparison. To generate these figures we used the following
parameters of signal, noise and interference. Pulse repetition
frequency was set to 20 kHz, interference consisted of two
components with average Dopplers 0 and 1000 Hz and Doppler spread
equal to 500 Hz, interference spectral envelope was Gaussian and
input signal to white noise ratio was 10 dB. Doppler shift of the
signal was 4 kHz, amplitude distribution of useful signal in
training sample was Rayleigh and average power of signal in training
sample (if present) was equal to its power in the cell of interest.
The number of iterations $T$ in proposed algorithm was 3. The
results were averaged over 500 Monte--Carlo trials.

Figure~\ref{fig:reg_aut_1} shows output SINR of LSMI algorithm when
training sample is not contaminated by the useful signal and
Fig.~\ref{fig:reg_aut_2} shows the results of training sample
contamination. By observing these figures we can conclude that when
useful signal is present in training sample, severe degradation of
LSMI performance occurs if fixed loading factor is used. The
algorithm using adaptive loading factor calculated using proposed
approach is able to alleviate this effect, especially if useful
signal dominates in the training sample. On the other hand, by
observing Fig.~\ref{fig:reg_aut_1} we can see that the introduction
of adaptive loading factor leads to insubstantial loss in
performance when useful signal is absent from the training sample.
Thus we can conclude that the application of the proposed approach
yields overall performance improvement when used in conjunction with
LSMI algorithm. We can explain this result by the fact that the
proposed algorithm tries to find a balance between the gain that is
obtained by applying matched filter for white noise case $\w = \s$
and the adaptive filter for colored noise case $\w =
\widehat\R^{-1}\s$. In the situation when useful signal is not
present in training sample, the application of the latter filter
gives the best gain. Whereas when the training sample is
contaminated by the useful signal, the application of this adaptive
filter may actually lead to significant performance degradation due
to signal cancellation effect. Thus in the latter situation the
proposed algorithm automatically increases the loading factor so
that more emphasis is put on non--adaptive weight vector $\w = \s$.
However, the observation of Fig.~\ref{fig:reg_aut_4:-60} and
Fig.~\ref{fig:reg_aut_4:-80} leads to the conclusion that this
loading factor tuning does not lead to significant performance
improvement when interference is very strong. This can be attributed
to the fact that in this case it is not possible to achieve
performance improvement by just varying the effect of the two
mentioned weight vectors.

\section{Concluding Remarks}
\label{section::Concluding Remarks}
In this paper an iterative algorithm for loading factor optimization
is proposed within the framework of LSMI algorithm. Linearization of
the expression for weights obtained via LSMI algorithm is used to
derive constrained empirical cost function. An expression for
loading factor minimizing this cost function is derived. After that,
an iterative solution is proposed to take into account the fact that
the linearization prevents the convergence of the proposed algorithm
during the first iteration. Finally, simulation examples showing the
effectiveness of the proposed scheme along with the discussion of
obtained results are presented. Analytical justification of the
algorithm described in this paper is the subject to further
development.

\bibliographystyle{IEEEtran}
\bibliography{abrev,jwsn}

\end{document}